\documentclass[twocolumn,showpacs,amsmath,amssymb,superscriptaddress,eqsecnum,prb]{revtex4}

\usepackage{graphicx}
\usepackage{dcolumn}
\usepackage{bm}

\newcommand{\aver}[1]{\langle #1 \rangle}

\begin{document}

\title{Theory of Transport through Quantum-Dot Spin Valves in the Weak-Coupling Regime}

\author{Matthias Braun}
\affiliation{Institut f\"ur Theoretische Physik III, Ruhr-Universit\"at Bochum, 44780 Bochum, Germany}
\affiliation{Institut f\"ur Theoretische Festk\"orperphysik, Universit\"at Karlsruhe, 76128 Karlsruhe, Germany}
\author{J\"urgen K\"onig}
\affiliation{Institut f\"ur Theoretische Physik III, Ruhr-Universit\"at Bochum, 44780 Bochum, Germany}
\affiliation{Institut f\"ur Theoretische Festk\"orperphysik, Universit\"at Karlsruhe, 76128 Karlsruhe, Germany}
\author{Jan Martinek}
\affiliation{Institut f\"ur Theoretische Festk\"orperphysik, Universit\"at Karlsruhe, 76128 Karlsruhe, Germany}
\affiliation{Institute for Materials Research, Tohoku University, Sendai 980-8577, Japan}
\affiliation{Institute of Molecular Physics, Polish Academy of Science, 60-179 Pozna\'n, Poland}

\date{\today}

\begin{abstract}
We develop a theory of electron transport through quantum dots that are
weakly coupled to ferromagnetic leads.
The theory covers both the linear and nonlinear transport regime, takes
non-collinear magnetization of the leads into account, and allows for an
externally applied magnetic field.
We derive generalized rate equations for the dot's occupation and accumulated
spin and discuss the influence of the dot's spin on the transmission.
A negative differential conductance and a nontrivial dependence of
the conductance on the angle between the lead magnetizations are predicted.
\end{abstract}

\maketitle

\section{\label{sec:level1}Introduction}
The study of spin-dependent transport phenomena has recently
attracted much interest and led to new device functionalities
based on the manipulation of the spin degree of
freedom.\cite{reviews,datta} A prominent example, which has
already proven technological relevance, is the spin valve based on
either the giant magnetoresistance effect\cite{GMR} in magnetic
multilayers or the tunnel magnetoresistance\cite{julliere} in
magnetic tunnel junctions. In the latter case, the tunneling
current between two ferromagnets depends on the relative
orientation of the leads' magnetizations. The maximal and minimal
transmission is achieved for the parallel and antiparallel
configurations, respectively. At intermediate angles $\phi$ of the
relative magnetization direction, the transmission is interpolated
by a cosine law. This angular dependence, which has been
beautifully demonstrated experimentally,\cite{angular} can be
easily understood within a noninteracting-electron
picture.\cite{slonczewski} It simply reflects the $\phi$-dependent
overlap of the spinor part of the majority-spin-electron wave
functions in the source and drain electrodes.

\begin{figure}[!ht]
\includegraphics[width=0.7\columnwidth,angle=0]{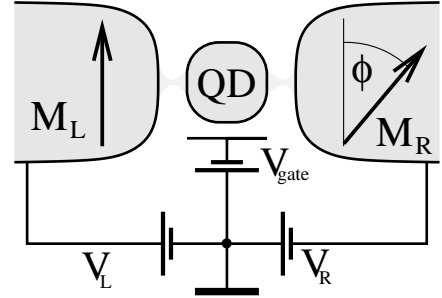}
\caption{\label{intro}
  Quantum-dot spin valve: a quantum dot is weakly coupled to ferromagnetic
  leads. The magnetization directions of the leads enclose an angle $\phi$.}
\end{figure}

On the other hand, tunneling transport through nanostructures, such as
semiconductor quantum dots or metallic single-electron transistors, is
strongly affected by Coulomb interaction, and a noninteracting-electron picture
is no longer applicable.
Recent works on spin-dependent tunneling through these devices include
studies on metallic islands,\cite{ono,metal-theory} granular systems,\cite{granular} and carbon nanotubes\cite{cnt} coupled to ferromagnetic
leads as well as spin-polarized transport from
ferromagnets through quantum dots.\cite{noise,QD-theory,QD-theory2,Kondo,QD-exp,linearresponse,angular2,ralph,ssplit}
The focus of this paper is on quantum-dot spin valves, i.e., quantum
dots attached to ferromagnetic leads, with non-collinear magnetizations,
see Fig.~\ref{intro}.

The much richer transport behavior of quantum-dot spin valves, as compared to
single magnetic tunnel junctions, relies on the possibility to generate a
non-equilibrium spin accumulation on the quantum dot, depending on system
parameters such as gate and bias voltage, charging energy, asymmetry of
the tunnel couplings, and external magnetic field.
This opens the potential of a controlled manipulation of the quantum-dot spin,
detectable in transport.
To some degree, there is a relation between the system under consideration
and a single magnetic-atom spin on a scanning tunneling microscope tip.
For the latter, precession of the single spin in an external magnetic field
has been detected in the power spectrum of the tunneling
current.\cite{precession_e,precession_t}

In a recent Letter by two of us,\cite{linearresponse} we predicted
that the interplay of spin-dependent tunneling and Coulomb
interaction in quantum-dot spin valves gives rise to an
interaction-driven spin precession, describable in terms of an
exchange field and detectable in a nontrivial
$\phi$ dependence of the linear conductance.
The aim of the present paper is to extend this discussion to a complete
theory of transport through quantum-dot spin valves in the limit
of weak dot-lead coupling.\cite{remark} The theory covers both the
regimes of linear and nonlinear response, and is easily
generalized to include an external magnetic field and/or spin
relaxation processes within the quantum dot. The quantities of
central interest are the magnitude and direction of the spin
accumulated on the quantum dot. They can be determined by
generalized kinetic equations within a diagrammatic real-time
transport theory for the dot's density matrix. To obtain a more
intuitive and easily accessible framework for the dynamics of the
dot's occupation and spin, however, we transform the kinetic
equations into rate equations that exhibit similarities to Bloch
equations of a spin in a magnetic field. The derivation of these
Bloch-like rate equations is the first central result of this
paper, presented in Sec.~\ref{sec:calculus} after defining the
model in Sec.~\ref{sec:model}. They provide a convenient starting
point for the discussion of how the dot state depends on the system parameters.
The second major achievement of this paper is to relate the current through the
device to the dot's occupation and average spin, which is done in
Sec.~\ref{sec:current}.

An analysis of the obtained results is presented in
Sec.~\ref{sec:discussion}.
We find that the quantum-dot spin, and thus the current, behaves quite
differently in the linear- and nonlinear-response regime.
In linear response, the above-mentioned effective exchange field
rotates the quantum-dot spin such that the spin-valve effect is reduced.
This contrasts with the nonlinear-response regime, in which the dot spin
tends to align antiparallel to the drain electrode, leading to a spin blockade.
Together with the effective exchange field this leads to a very pronounced
negative differential conductance.

We, then, conclude by generalizing our theory to include an
external magnetic field (Sec.~\ref{sec:magnetic_field}) or
internal spin relaxation in the dot
(Sec.~\ref{sec:spin_relaxation}), and by interpreting the
effective exchange field within the recently introduced\cite{spinmix}
language of the spin-mixing conductance (Sec.~\ref{sec:mixing_conductance}).

\section{Model}
\label{sec:model}

We consider a quantum dot with a level spacing exceeding thermal broadening,
intrinsic linewidth, and charging energy, i.e., only a single level
with energy $\varepsilon$, measured relative to the Fermi energy of the leads,
contributes to transport.
The quantum dot is tunnel-coupled to ferromagnetic leads.
We model this system with the Hamiltonian
\begin{eqnarray}
  H=H_{\rm dot}+H_{\rm L}+H_{\rm R}+H_{\rm T,L}+H_{\rm T,R} \, .
\end{eqnarray}
The first part $H_{\rm dot}=\sum_{\sigma}\varepsilon
c^{\dag}_{\sigma}c_{\sigma} + U\,n_{\uparrow}n_{\downarrow}$
describes an atomiclike spin-degenerate energy level on the dot
plus the charging energy $U$ for double occupancy. The left and
right ferromagnetic leads, $r={\rm L,R}$, are treated as a
reservoir of itinerant electrons, $H_r = \sum_{k\sigma}
\varepsilon^{\,}_{k\sigma} a^{\dag}_{rk\sigma}a^{\,}_{rk\sigma}$,
where $\sigma = +$ labels majority and $\sigma = -$ minority-spin
electrons. Here, the quantization axis for the electron spins in
reservoir $r$ is chosen along its magnetization direction ${\bf
\hat{n}}_{r}$. In the spirit of a Stoner model of ferromagnetism,
we assume a strong spin asymmetry in the density of states
$\rho_{r,\pm}(\omega)$ for majority ($+$) and minority ($-$) spins.
(To be more precise, without loss of generality we assume the direction
of magnetization to be parallel to the direction of majority spins.)
In the following, the densities of states are approximated
to be energy independent $\rho_{r,\pm}(\omega) = \rho_{r,\pm}$.
(Real ferromagnets have a structured density of states. This will modify
details of our results but not change the general physical picture.)
The asymmetry in the density of states is characterized by the
degree of spin polarization
$p_r=(\rho_{r+}-\rho_{r-})/(\rho_{r+}+\rho_{r-})$ with $0\le p_r
\le 1$, where $p_r=0$ corresponds to a nonmagnetic lead, and
$p_r=1$ describes a half-metallic lead, carrying majority
spins only. The magnetization directions of the leads can differ
from each other, enclosing an angle $\phi = \sphericalangle
({\bf\hat{n}}_{\rm L},{\bf\hat{n}}_{\rm R})$. An applied bias
voltage is accounted for in different electrochemical potentials
for the left and right leads.

\begin{figure}[!ht]
\includegraphics[width=0.95\columnwidth,angle=0]{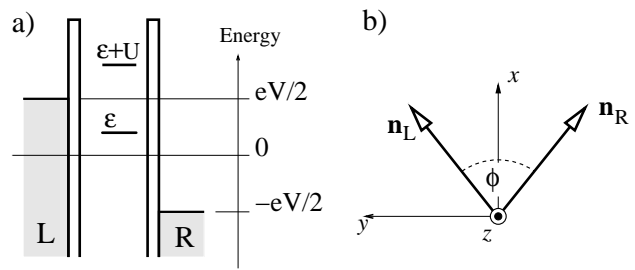}
\caption{
 \label{reference_frame}
 (a) Scheme of the quantum-dot energy scales.
 The difference of the electrochemical potentials of the left and
 right leads describes a symmetrically applied bias voltage.
 (b) The coordinate system we choose.
 The magnetization directions ${\bf \hat{n}}_{\rm L}$ and
 ${\bf \hat{n}}_{\rm R}$ enclose an angle $\phi$.}
\end{figure}

Tunneling between leads and dot is described by tunnel Hamiltonians
$H_{\rm T,L}$ and $H_{\rm T,R}$.
Their explicit form depends on the choice of the spin quantization axis for
the dot states.
We find it convenient to choose neither ${\bf \hat{n}}_{\rm L}$ nor
${\bf \hat{n}}_{\rm R}$ but instead to quantize the dot spin
$\sigma = \uparrow, \downarrow$ along the $z$-direction of the coordinate
system in which the basis vectors ${\bf \hat{e}}_{\rm x}$,
${\bf \hat{e}}_{\rm y}$, and ${\bf \hat{e}}_{\rm z}$ are along
${\bf \hat{n}}_{\rm L} + {\bf \hat{n}}_{\rm R}$,
${\bf \hat{n}}_{\rm L} - {\bf \hat{n}}_{\rm R}$, and
${\bf \hat{n}}_{\rm R} \times {\bf \hat{n}}_{\rm L}$, respectively
(see also Fig.~\ref{reference_frame}).
The tunnel Hamiltonian for the left tunneling barrier then reads
\begin{eqnarray}
\label{tp}
H_{\rm T,L}=\frac{t_{\rm L}}{\sqrt{2}}\sum_k
\!\!\!\!\!\! &  & a^{\dag}_{{\rm L}k+}\bigl( e^{+i\phi/4}c^{}_{\uparrow}+e^{-i\phi/4}c^{}_{\downarrow}\bigl)\nonumber\\
+\!\!\!\!\!\!&  & a^{\dag}_{{\rm L}k-}\bigl(-e^{+i\phi/4}c^{}_{\uparrow}+e^{-i\phi/4}c^{}_{\downarrow}\bigl)\,
+\,\text{H.c.} \qquad
\end{eqnarray}
and for the right barrier, $H_{\rm T,R}$, the same with the
replacements ${\rm L} \rightarrow {\rm R}$ and $\phi \rightarrow
-\phi$. Here, $t_{\rm L}$ and $t_{\rm R}$ are the tunnel matrix
elements. The tunneling rate for electrons from lead $r$ with spin
$(\pm)$ is then $\Gamma_{r,\pm}/\hbar = 2\pi|t_r|^2 \rho_{r,\pm}/\hbar$.
Therefore the electronic states in the quantum dot with spin
$(\uparrow\downarrow)$ acquire a finite line width
$\Gamma_r=\sum_{\sigma=\pm}2\pi|t_r/\sqrt{2}|^2 \rho_{r,\sigma}=
\sum_{\sigma=\pm}\Gamma_{r,\sigma}/2$.

\begin{figure}[!ht]
\includegraphics[width=0.8\columnwidth,angle=0]{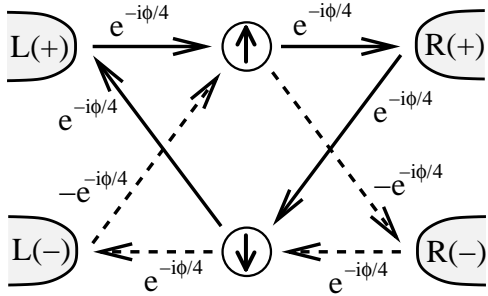}
\caption{
  \label{tunnelphases}
  Graphical representation of the phase factors in the tunnel Hamiltonian,
  Eq.~(\ref{tp}).
  In this representation, the quantum-dot spin valve bears similarities to an
  Aharonov-Bohm setup, with the different arms modeling the two different
  spin channels $\uparrow$ and $\downarrow$, and the Aharonov-Bohm flux
  corresponding to the relative angle of the leads' magnetizations.}
\end{figure}

We note that while
individual parts of the tunnel Hamiltonians do not conserve spin
separately, the sum of all parts strictly does. Due to the special
choice of the quantization axis, the lead electrons are coupled
equally strong to up- and down-spin states in the quantum dot.
There are, however, phase factors involved, similar to
Aharonov-Bohm phases in multiply connected geometries. The formal
similarity of the quantum-dot spin valve to a two-dot
Aharonov-Bohm interferometer\cite{AB2} is visualized in
Fig.~\ref{tunnelphases}.

\section{Occupation and spin of the dot}
\label{sec:calculus}

In this section we derive rate equations governing the state of the dot,
i.e., its occupation as well as the magnitude and direction of an
accumulated spin.
For this, we make use of a diagrammatic real-time approach\cite{diagrams} to
describe the time evolution of the reduced density matrix for the dot
degree of freedoms, which is obtained after integrating out the lead electrons.
The resulting equations for the density matrix elements can be reformulated
into a more intuitive language of rate equations for the dot's occupation and
spin.

\subsection{Reduced density matrix for the quantum dot}

The dynamics of the system can be described by the time evolution
of the total density matrix, which accounts for all degrees of freedom
in the dot and the leads.
While the leads are treated as reservoirs of noninteracting electrons,
it is the dynamics of the dot degrees of freedom that determine the transport
behavior.
This motivates the idea to integrate out the leads and arrive at a reduced
density matrix for the dot degrees of freedom only.
Details about the procedure of integrating out the leads are given in
Ref.~\onlinecite{diagrams}.

The basis of the Hilbert space is given by the states $\chi=0$ (empty dot), $\uparrow$ (dot occupied
with one spin-up electron), $\downarrow$ (the same with a spin-down electron),
and ${\rm d}$ (double occupancy of the dot).
Therefore, the reduced density matrix $\rho_{\rm dot}$ is a $4 \times 4$
matrix with matrix elements
$P_{\chi_2^{}}^{\chi_1} \equiv (\rho_{\rm dot})_{\chi_1 \chi_2}$.
We get
\begin{eqnarray}
  \label{dotdm}
  \rho_{\rm dot}= \left(
  \begin{array}{cccc}
    P_0 &     0 &     0 &     0 \\
    0     & P_{\uparrow} & P^{\uparrow}_{\downarrow} &     0 \\
    0     & P^{\downarrow}_{\uparrow} & P_{\downarrow} &     0 \\
    0     &     0 &     0 & P_{\rm d}
  \end{array} \right) \, .
\end{eqnarray}
The diagonal, real entries $P_\chi \equiv P_\chi^\chi$ are nothing
but the probabilities to find the dot in the corresponding state
$\chi = 0,\uparrow, \downarrow, {\rm d}$. The zeros in
Eq.~(\ref{dotdm}) in the off diagonals are a consequence of the
total-particle-number conservation. The nonvanishing complex
off-diagonal elements $P_{\downarrow}^{\uparrow} =
(P_{\uparrow}^{\downarrow})^*$ reflect the fact that an
accumulated dot spin is not restricted to the $z$ direction but
can also have a finite $x$ and $y$ component. The density matrix
has, thus, six real parameters (four diagonal elements and the
real and imaginary parts of $P_{\uparrow}^{\downarrow}$). Due to
the normalization condition ${\rm Tr} \, \rho_{\rm dot} = 1$,
only five of them are independent.

The time evolution of the reduced density matrix is given by a
generalized master equation in Liouville space\cite{diagrams},
\begin{eqnarray}
  \label{master}
  \frac{d}{dt} P_{\chi_2^{}}^{\chi_1}(t) &=&
  -\frac{i}{\hbar}(E_{\chi_1}-E_{\chi_2})P_{\chi_2^{}}^{\chi_1}(t)
  \nonumber \\
  &&+\int\limits_{-\infty}^t dt^{\prime}\,
  \sum_{\chi_1^{\prime} \chi_2^{\prime}}
  {P_{\chi_2^{\prime}}^{\chi_1^{\prime}}}(t^{\prime}) \,\,
  {\Sigma_{\chi_2^{\prime}\,\chi_2}^{\chi_1^{\prime}\,\chi_1}}(t^{\prime},t)
  \, . \qquad
\end{eqnarray}
Here, $E_{\chi_i^{\,}}$ are the energies of the decoupled system
(in our case $E_0 = 0$, $E_\uparrow = E_\downarrow = \varepsilon$,
and $E_{\rm d} = 2\varepsilon + U$). The kernels
$\Sigma_{\chi_2^{\prime}\,\chi_2}^{\chi_1^{\prime}\,\chi_1}
(t^{\prime},t)$ act as generalized transition rates in Liouville
space. They are defined as irreducible self-energy parts of the
dot propagator on a Keldysh contour (see appendix
\ref{appendix_sigmas}), and can be expanded in powers of
the dot-lead coupling strength $\Gamma$. In the limit of interest
for this paper, i.e., weak dot-lead coupling, only the terms
linear in $\Gamma$ are retained.

The Markovian approximation is established by replacing
$P_{\chi_2^{\prime}}^{\chi_1^{\prime}}(t^{\prime})$ in Eq.~(\ref{master}) with
$P_{\chi_2^{\prime}}^{\chi_1^{\prime}}(t)$.
In this case, the time integral can be comprised in the definition
$\Sigma_{\chi_2^{\prime}\,\chi_2}^{\chi_1^{\prime}\,\chi_1} =
i\hbar \int_{-\infty}^0 dt^{\prime}
\Sigma_{\chi_2^{\prime}\,\chi_2}^{\chi_1^{\prime}\,\chi_1}(t^{\prime},0)$.
We will only discuss results for the steady-state limit, i.e.,
all matrix elements are time independent and the above-mentioned replacement
is not an approximation but valid in general.
In order to emphasize the physical origin of the rate equations, however, we
will keep all time derivatives although their numerical value is zero for our
discussion.

\subsection{Master and Bloch-like equations}
For a more intuitive access to the average dot state, we connect the matrix
elements of the reduced density matrix to the vector of the average spin
$\hbar {\bm S}$ with ${\bm S} = (S_{\rm x},S_{\rm y},S_{\rm z})$ by
\begin{eqnarray}
  \label{spinreplacement}
  S_{\rm x} = \frac{P^{\uparrow}_{\downarrow}+P^{\downarrow}_{\uparrow}}{2}
  ;\;\;
  S_{\rm y} = i\frac{P^{\uparrow}_{\downarrow}-P^{\downarrow}_{\uparrow}}{2}
  ;\;\;
  S_{\rm z} = \frac{P_{\uparrow}-P_{\downarrow}}{2} .\quad
\end{eqnarray}
We emphasize that the ${\bm S}$ describes a quantum-statistical average rather
than the coherent state of a single spin.
As a consequence, the magnitude $|{\bm S}|$ can adopt any number between
$0$ and $1/2$.
The average charge occupation of the dot is determined by the probabilities
$P_0$, $P_1$, and $P_{\rm d}$, where $P_1=P_{\uparrow}+P_{\downarrow}$.
The dot state is, thus, characterized by the set of parameters
$(P_0,P_1,P_{\rm d},S_{\rm x},S_{\rm y},S_{\rm z})$, in which only five of them
are independent, since the normalization of the probabilities requires
$P_0 + P_1 + P_{\rm d}=1$.

We evaluate the irreducible self-energies appearing in Eq.~(\ref{master}) up
to first order in $\Gamma$ (see Appendix~\ref{diagramtechnique}) to get a
set of six master equations, three for the occupation probabilities and three
for the average spin.
The first three are given by
\begin{widetext}
\begin{eqnarray}
\label{P01d}
  \frac{d}{dt} \left(
  \begin{array}{c}
    P_0 \\ P_1 \\ P_{\rm d}
  \end{array}
  \right) &=& \sum_{r=\rm L,R} \frac{\Gamma_r}{\hbar} \left(
  \begin{array}{ccc}
    - 2f^+_r(\varepsilon) & f^-_r(\varepsilon) & 0 \\
    \hphantom{-}2f^+_r(\varepsilon) & - f^-_r(\varepsilon) - f^+_r(\varepsilon+U)
    & \hphantom{-}2f^-_r(\varepsilon+U) \\
    0 & f^+_r(\varepsilon+U) & - 2f^-_r(\varepsilon+U)
  \end{array}
  \right) \left(
  \begin{array}{c}
    P_0 \\ P_1 \\ P_{\rm d}
  \end{array}
  \right)
\nonumber \\
  &&
  + \sum_{r=\rm L,R} \frac{2 p_r \Gamma_r}{\hbar} \left(
  \begin{array}{c}
    f^-_r(\varepsilon) \\
    - f^-_r(\varepsilon) + f^+_r(\varepsilon+U) \\
    -f^+_r(\varepsilon+U)
  \end{array}
  \right)
    {\bm S}\cdot {\bf \hat n}_r \, .
\end{eqnarray}

\end{widetext}
Here, we have used the following definitions.
The probability to find an electronic state at energy $\omega$ in the left
(right) lead occupied is given by the Fermi function $f_{\rm L(R)}^+(\omega)$
and by $f_{\rm L(R)}^-(\omega) = 1-f_{\rm L(R)}^+(\omega)$ to find it empty.

In the case of nonmagnetic leads, $p_r=0$, only the first line in
Eq.~(\ref{P01d}) survives, and we recover the usual master
equations for sequential tunneling through a single level.\cite{rateargument}
For spin-polarized leads, however, tunneling into or out of the dot becomes
spin dependent.
As a consequence, the charge occupation, in general, depends on the spin
${\bm S}$ accumulated on the dot, as indicated by the second line of
Eq.~(\ref{P01d}).
The three remaining master equations describe the time evolution of the
average spin,
\begin{equation}
\label{S}
  \frac{d\bm S}{dt} = \left( \frac{d\bm S}{dt}\right)_{\rm acc}
  + \left( \frac{d\bm S}{dt}\right)_{\rm rel}
  + \left( \frac{d\bm S}{dt}\right)_{\rm rot}
\end{equation}
with
\begin{widetext}
\begin{eqnarray}
\label{Sacc}
  \left( \frac{d\bm S}{dt}\right)_{\rm acc} &=& \sum_{r=\rm L,R}
  \frac{p_r \,\Gamma_r}{\hbar}
  \left[
    f^+_r(\varepsilon) P_0
    + \frac{ - f^-_r(\varepsilon) + f^+_r(\varepsilon+U)}{2} P_1
    - f^-_r(\varepsilon+U) P_{\rm d}
  \right] {\bf \hat n}_r
\\
\label{Srel}
\left( \frac{d\bm S}{dt}\right)_{\rm rel}
&=&-\sum_{r=\rm L,R}\frac{\Gamma_r}{\hbar}\left[
  f^-_r(\varepsilon) + f^+_r(\varepsilon+U) \right] {\bm S}
\\
\label{Srot}
\left( \frac{d\bm S}{dt}\right)_{\rm rot} &=&
     {\bm S} \times \sum_{r=\rm L,R} {\bm B}_r \, ,
\end{eqnarray}
\end{widetext}
where we used the definition
\begin{equation}
\label{exchange}
  {\bm B}_r = p_r \,\frac{\Gamma_r{\bf \hat n}_r}{\pi\hbar} \int'
  d\omega \left( \frac{f^+_r(\omega)}{\omega-\varepsilon-U}
    +\frac{f^-_r(\omega)}{\omega-\varepsilon} \right)
\end{equation}
for $r={\rm L},{\rm R}$, and the prime at the integral symbolizes
Cauchy's principal value. We see that three different
contributions lead to a change of the average dot spin. The first
one, Eq.~(\ref{Sacc}), depends on the occupation
probabilities only and describes nonequilibrium spin accumulation
via tunneling to and from spin-polarized leads. This is the source
term, which is responsible for building up a spin polarization in
the quantum dot. The second term (proportional to $\bm S$) has the
opposite effect. It accounts for the decay of the dot spin by tunneling out
of the electron with given spin or by tunneling in of a
second electron with opposite spin forming a spin singlet on
the dot. These tunneling mechanisms yield an intrinsic spin
relaxation rate $1/ \tau_{\rm c} = \sum_{r=\rm L,R}\, (\Gamma_r/\hbar)
\left[ f^-_r(\varepsilon) + f^+_r(\varepsilon+U) \right]$. From Eq. (\ref{S})
we can derive the magnitude of the spin on the dot via
\begin{equation}
\label{Smag}
  \frac{d |{\bm S}|}{dt} = \frac{\bm S}{|{\bm S}|} \cdot
  \left( \frac{d\bm S}{dt}\right)_{\rm acc} - \frac{|{\bm S}|}{\tau_{\rm c}} \, .
\end{equation}

The two contributions, Eqs.~(\ref{Sacc}) and (\ref{Srel}), capture
the spin dynamics due to spin-dependent tunneling processes
transferring electrons into and out of the quantum dot.
But above equation is also sensitive to the direction of $\bm S$,
which depends directly on the third contribution, Eq.~(\ref{Srot}),
of the Bloch-like equation. This term leads to a
precession of the dot spin about ${\bm B}_{\rm L} + {\bm B}_{\rm
R}$, where $({\bm B}_{\rm L} + {\bm B}_{\rm R})/\gamma$ can be
viewed as a fictitious magnetic field, with $\gamma=-g \mu_{\mathrm B}$
being the gyromagnetic ratio. This fictitious field vanishes in the absence
of Coulomb interaction, $U=0$, i.e., its origin is a many-body
interaction effect. Its nature can be conceived as an
exchange field due to virtual particle exchange with the
spin-polarized leads. Note that these virtual exchange processes
do not change the charge of the dot, in contrast to the
spin-dependent tunneling events responsible for the first two
contributions to Eq.~(\ref{S}). Nevertheless, this contribution is
also linear in $\Gamma$ and has to be kept to be consistent. We
emphasize that the use of a simple rate-equation picture, which
neglects this exchange field and takes into account spin-dependent
tunneling only, would be fundamentally insufficient.

The type of exchange interaction that emerges in our theory has
been discussed in the literature in the context of Kondo physics
for magnetic impurities in (normal) metals.\cite{fermions} With
the help of a Schrieffer-Wolff transformation, the Anderson
Hamiltonian describing the magnetic impurity can be transformed to
an s-d model, in which the spin of the magnetic impurity is
coupled to the conduction-band-electron spins of the metal. Since
the metal is spin symmetric, there is no net exchange field,
though. This contrasts with our model, which includes a finite
spin polarization of the leads. By integrating out the lead
electrons of the transformed Hamiltonian in the subspace
of single dot occupancy, we recover the precise mathematical form
of the exchange field as given in Eq.~(\ref{exchange}).

The generalization of our theory to the case of real ferromagnets
with nonuniform density of states is straightforward. The only
modification is to keep the energy dependence of
$p_{ r}(\omega) \equiv
[\rho_{ r +}(\omega)-\rho_{ r-}(\omega)]/
[\rho_{ r +}(\omega)+\rho_{ r-}(\omega)]
$ and $\Gamma_{r}(\omega)$ in all formulas. In particular,
these quantities have to be included
in the energy integral for the exchange field in
Eq.~(\ref{exchange}).\cite{proc,NRG}

We remark that the effective exchange field is not only
responsible for the torque on an accumulated spin as discussed
above, but it also generates a spin splitting of the dot level.\cite{remark2} Since this Zeeman-like
splitting is proportional to $\Gamma$, it cannot be resolved in
any transport signal in the weak-coupling regime considered in the
present paper. This splitting gives rise to a correction of higher
order in the coupling that has to be dropped in a consistent
first-order-transport calculation. This contrasts with the limit
of strong lead-dot coupling, in which any spin splitting modifies
the emerging Kondo physics in a crucial manner.\cite{Kondo}

\section{Electric current}
\label{sec:current}
The current through the device can be expressed in terms of
Keldysh Green's functions of electrons, as shown, e.g., in
Refs.~\onlinecite{Meir,AB2}.
The starting point of our discussion is the relation for the current $I_r$
coming from reservoir $r$,
\begin{eqnarray}
\label{landauer}
  I_r = \frac{i(-e)}{h} \int d\omega \,\, {\rm tr}
  \left\{{ \bm \Gamma}_r^{} f_r^+(\omega){\bm G}^>(\omega)
    + {\bm \Gamma}_r^{} f_r^-(\omega){\bm G}^<(\omega) \right\} . \nonumber \\
\end{eqnarray}
Here, the bold face indicates a $2\times 2$ matrix structure for the spin
degree of freedom.
The lesser and greater Green's functions
${\bm G^<} (\omega) = \left(
\begin{array}{cc}
  G^<_{\uparrow\uparrow} (\omega) & G^<_{\uparrow\downarrow} (\omega)\\
  G^<_{\downarrow\uparrow} (\omega) & G^<_{\downarrow\downarrow} (\omega)
\end{array} \right)$
and ${\bm G^>} (\omega)$ are defined as the Fourier transforms of
$G^<_{\sigma \sigma'} (t) = i \langle c^\dagger_{\sigma'}(0)c_{\sigma}(t)
\rangle$ and
$G^>_{\sigma \sigma'} (t) = -i \langle c_{\sigma}(t)c^\dagger_{\sigma'}(0)
\rangle$, respectively.
The coupling matrices are defined as
${\bm \Gamma}_{\rm L} = \Gamma_{\rm L} \left(
\begin{array}{cc}
  1 & p_{\rm L}\,e^{- i\phi/2} \\
  p_{\rm L}\,e^{+ i\phi/2} & 1 \\
\end{array} \right)$,
and the same for ${\bm \Gamma}_{\rm R}$ but with the substitutions
${\rm L} \rightarrow {\rm R}$ and $\phi \rightarrow -\phi$.
The diagonal matrix elements describe tunneling of spins along the
$z$ direction and are, therefore, independent of the leads' spin polarization
$p_{r}$ since the magnetizations of the leads lie in the $x-y$ plane.
This contrasts with the off-diagonal matrix elements taking into account
tunneling of spins in the $x-y$ plane.

In the weak-coupling limit, the current is dominated by the contributions of
first order in the coupling $\Gamma$.
Because $\Gamma$ already explicitly appears in Eq.~(\ref{landauer}), the
Green's functions have to be calculated to zeroth order in
$\Gamma$.\cite{footnote}
The final result for the current can be written as a sum over two contributions,
\begin{equation}
\label{current}
  I_r = I_r^{\rm P} + I_r^{\rm S} \, ,
\end{equation}
which explicitly contain either the occupation
probabilities $P_0,P_1,P_{\rm d}$ or the accumulated spin ${\bm S}$,
respectively,
\begin{eqnarray}
\label{currentP}
    I_r^{\rm P} &=& \Gamma_{r}\frac{2(-e)}{\hbar} \left[
    f^+_{r}(\varepsilon) P_0 +
    \frac{f^+_{r}(\varepsilon+U)-f^-_{r}(\varepsilon)}{2} P_1 \qquad
\right. \nonumber \\ && \left. \qquad\,\,\,\,\,\,
    - f^-_{r}(\varepsilon+U) P_{\rm d}   \vphantom{\frac{f^+_{r}(\varepsilon+U)-f^-_{r}(\varepsilon)}{2}}\right],
\\
  I_r^{\rm S} &=& -p_r\,\Gamma_r\,\frac{2(-e)}{\hbar}  \,
  \left[ f^-_r(\varepsilon) + f^+_r(\varepsilon+U) \right]
    {\bm S}\cdot {\bf \hat n}_r \, .
\label{currentS}
\end{eqnarray}
For nonmagnetic leads, only the term $I_r^{\rm P}$ contributes.
At finite spin polarization in the leads, the additional term $I_r^{\rm S}$
appears, but also the first contribution $I_r^{\rm P}$ is affected via the
modification of the occupation probabilities as given in Eq.~(\ref{P01d}).

In steady state the two currents $I_{\rm L}$ and $I_{\rm R}$ are connected
by the conservation of charge on the dot, so the
homogeneous current flowing through the system can be defined as
$I=I_{\rm R}=-I_{\rm L}$.

\section{Results}
\label{sec:discussion}

In the following we discuss results for the spin accumulation and its
impact on transport for both the linear and nonlinear regime.
To be specific, we choose symmetric coupling
$\Gamma_{\rm L} = \Gamma_{\rm R} = \Gamma/2$, equal spin polarizations
$p_{\rm L} = p_{\rm R} = p$, and a symmetrically applied bias
$V_{\rm R} = -V_{\rm L} =  V/2$.

\subsection{Linear response}
\label{sec:linearresponse}
In equilibrium, i.e., without any applied bias voltage, $V=0$, no current
flows and the stationary solution of the rate equations
for the occupation probabilities (\ref{P01d}) is given by the Boltzmann
distribution, $P_\chi \sim \exp(- E_\chi/k_{\rm B}T)$, and since in this case
$(d{\bm S}/dt)_{\rm acc} = 0$, no spin accumulation occurs in the quantum
dot, ${\bm S} =0$.\cite{com1}

At nonzero bias voltage, a current flows through the system and spin is
accumulated on the dot. To describe the dot
state in the linear-response regime, $eV \ll  k_{\rm B} T$, we
expand the steady-state solution of Eqs.~(\ref{P01d}) and
(\ref{S}) up to linear order in $V$.
We find that (due to symmetric coupling) the dot occupation probabilities
$(P_0,P_1,P_{\rm d})$ have no corrections linear in $V$ and are, thus,
unchanged, independent of the size and direction of the leads magnetization.
In contrast to the occupation probabilities, the source term for the spin
accumulation has a nonzero contribution linear in $V$,
\begin{eqnarray}
    \left( \frac{d\bm S}{dt}\right)_{\rm acc} \!\!\!\!\!&=&\!
    \left[ f'(\varepsilon) \left( P_0 + \frac{P_1}{2} \right)
    + f'(\varepsilon+U) \left( \frac{P_1}{2} + P_{\rm d} \right) \right]
\nonumber \\
    && \times \frac{p\Gamma}{\hbar} \,\frac{-eV}{2k_{\rm B}T}
     ({\bf \hat n}_{\rm L} - {\bf \hat n}_{\rm R})
\, ,
\end{eqnarray}
generating a finite spin polarization along ${\bf \hat n}_{\rm L} - {\bf \hat n}_{\rm R}$,
i.e., along the $y$ axis.
This is consistent with the notion of angular momentum
conservation when magnetic moments polarized along
${\bf \hat n}_{\rm L}$ enter from the left than those along
${\bf \hat n}_{\rm R}$ leave the dot to the right.

\begin{figure}[!ht]
\begin{center}
\includegraphics[width=0.8\columnwidth,angle=0]{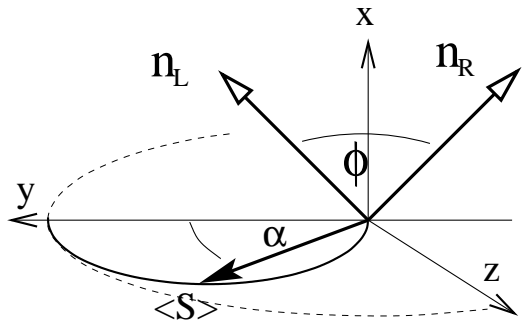}
\caption{
\label{linearresponse}
   Spin dynamics in the linear-response regime. Spin accumulates along
   the $y$ direction. The spin precesses due to the exchange field that is
   along the $x$ direction.
   Therefore, the stationary solution of the average spin on the dot is tilted
   away from the $y$ axis by an angle $\alpha$, plotted
   in Fig.~\ref{linrescurrentplots}(b).}
\end{center}
\end{figure}

The damping term, $(d{\bm S}/dt)_{\rm rel} = - (\Gamma/\hbar) \left[
f^-(\varepsilon) + f^+(\varepsilon+U) \right] {\bm S}$, limits the magnitude of
spin accumulation, but it does not affect its direction.
The third term, however, $(d{\bm S}/dt)_{\rm rot} = {\bm S} \times \left(
{\bm B}_{\rm L} + {\bm B}_{\rm R}\right)$ with ${\bm B}_r =  {\bf \hat n}_r p
\Gamma /(2\pi\hbar) \int' d\omega \left( \frac{1}{\omega-\varepsilon-U}
-\frac{1}{\omega-\varepsilon} \right) f(\omega)$ and
${\bm B}_{\rm L}+{\bm B}_{\rm R} \equiv B_0 \cos(\phi/2)\, \hat{\bf e}_x$, yields a precession of the
spin with an angle
\begin{equation}
  \alpha = - \arctan \left( B_0 \tau_{\rm c} \cos \frac{\phi}{2} \right)
\end{equation}
about ${\bf \hat n}_{\rm L} + {\bf \hat n}_{\rm R}$, i.e., the $x$-axis.
As a result, the accumulated spin acquires both $y$ and $z$ components as
seen in Fig.~\ref{linearresponse}.
This precession acts back on the magnitude of the accumulated spin.
The stationary solution of Eq.~(\ref{Smag}) yields
$|{\bm S}| = \tau_{\rm c} \left| \left( d {\bm S}/dt \right)_{\rm acc} \right|
\cos \alpha$, from which we infer that the precession is accompanied with
a decrease of the spin accumulation.

\begin{figure}[!ht]
\begin{center}
\includegraphics[width=1.0\columnwidth,angle=0]{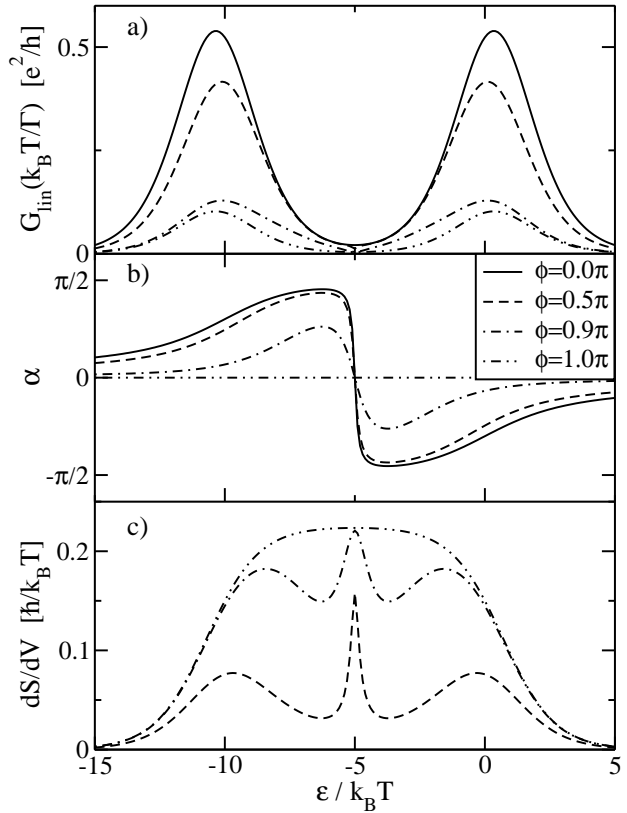}
\caption{\label{linrescurrentplots} (a) Linear
conductance normalized by $\Gamma/k_{\rm B}T$ as a function of the level
position $\varepsilon$ for different angles $\phi$. (b) Angle $\alpha$ enclosed by the
accumulated spin and the $y$ axis as defined in
Fig.~\ref{linearresponse}. (c) Derivation of the magnitude of the accumulated spin on the dot with respect to the
source-drain voltage $V$. Further parameters are $p=0.9$ and $U=10k_{\rm B}T$.}
\end{center}
\end{figure}

The rotation angle $\alpha$ is plotted in Fig.~\ref{linrescurrentplots}(b) as a
function of the level position $\varepsilon$.
We see that $\alpha$ changes sign at $\varepsilon = -U/2$, which reflects a
sign change of the exchange field at this point.
The point $\varepsilon = -U/2$ is special, as in this case the system bears
particle-hole symmetry.
The fast variation of $\alpha$ in this region gives rise to a rather sharp
feature in the magnitude of accumulated spin as a function of $\varepsilon$,
see Fig.~\ref{linrescurrentplots}(c).
The width of the emerging peak scales (for $k_{\rm B}T<U/2$) with
$\pi U [f^-(\varepsilon) + f^+(\varepsilon+U)]/[p\, \cos(\phi/2)]$, which
can be even smaller than temperature since the factor with the Fermi functions
involved can become numerically small.

\begin{figure}[!ht]
\begin{center}
\includegraphics[angle=-90,width=0.9\columnwidth]{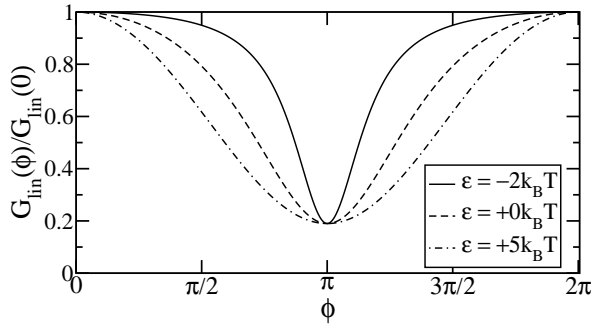}
\caption{\label{normalizedconductance} Normalized conductance as
a function of the angle $\phi$ enclosed by the lead magnetization
for different level positions and the parameters $U=10k_{\rm B}T$ and $p=0.9$.}
\end{center}
\end{figure}

As pointed out above, in the regime under consideration the occupation
probabilities do not depend on the spin polarization of the leads.
In particular, they are independent of the relative angle $\phi$ of the
leads' magnetization.
This means that the $\phi$ dependence of the conductance will be completely
determined by the product
${\bm S \cdot}{\bf \hat n}_{\rm L} = - {\bm S \cdot}{\bf \hat n}_{\rm R}$,
as can be seen from Eqs.~(\ref{currentP}) and (\ref{currentS}).
It is the relative orientation of the accumulated spin and the drain (or
source) that produces the $\phi$ dependence of the current, rather than the
product ${\bf \hat n}_{\rm L} {\bf \cdot} {\bf \hat n}_{\rm R}$, as in the
case of a single magnetic tunnel junction.
In this way, the $\phi$ dependent linear conductance $G^{\rm lin} =
(\partial I/ \partial V)|_{V=0}$ directly reflects the accumulated spin.
The effect of the exchange field $\bf B$ is seen from the analytic expression
\begin{eqnarray}
\label{currlin}
   \frac{G^{\rm lin}(\phi)}{G^{\rm lin}(0)} &=& 1
   - p^2\frac{\sin^2 (\phi/2)}{1+ (B_0 \tau_{\rm c})^2 \cos^2(\phi/2)}
   \, , \qquad
\end{eqnarray}
which is plotted in Fig.~\ref{normalizedconductance} for different values
of the level position $\varepsilon$.

For $\varepsilon > 0$, the quantum dot is predominantly empty, and
for $\varepsilon+U < 0$ doubly occupied with a spin singlet.
This is reflected by a small lifetime $\tau_{\rm c}$ of a dot spin.
Due to the short dwell time of a single spin-polarized electron in the dot, the
rotation angle is small and the normalized conductance as a function
of the relative angle $\phi$ of the lead magnetizations shows a
harmonic behavior, see, e.g., the curve for $\varepsilon = 5 k_{\rm B}T$ in
Fig.~\ref{normalizedconductance}.

For $-U  < \varepsilon < 0$, however, the dwell time
is increased and the exchange field becomes important.
It causes the above described spin precession, which decreases the product
${\bm S \cdot}{\bf \hat n}_{\rm L}$ since the relative angle
$\sphericalangle({\bf\hat{n}_{\rm L}},{\bf S})$ is increased and the magnitude
of ${\bm S}$ is reduced.
According to Eq.~(\ref{currentS}), the spin precession, thus, makes the
spin-valve effect less pronounced, leading to a value of the conductance
that exceeds the expectations made by Slonczewski\cite{slonczewski}
for a single magnetic tunnel junction.

For parallel and antiparallel aligned lead magnetizations, $\phi=0$ and
$\phi=\pi$, the accumulated spin and the exchange field also get aligned.
As a result, there is no spin precession, even though the exchange field
is still present.
For these cases, the $\phi$-dependent conductance is not affected by the
exchange field, see Fig.~\ref{normalizedconductance}.

\subsection{Nonlinear response}
\label{subsec:non_lin_response}

\begin{figure}[!ht]
\begin{center}
\includegraphics[width=1.0\columnwidth,angle=0]{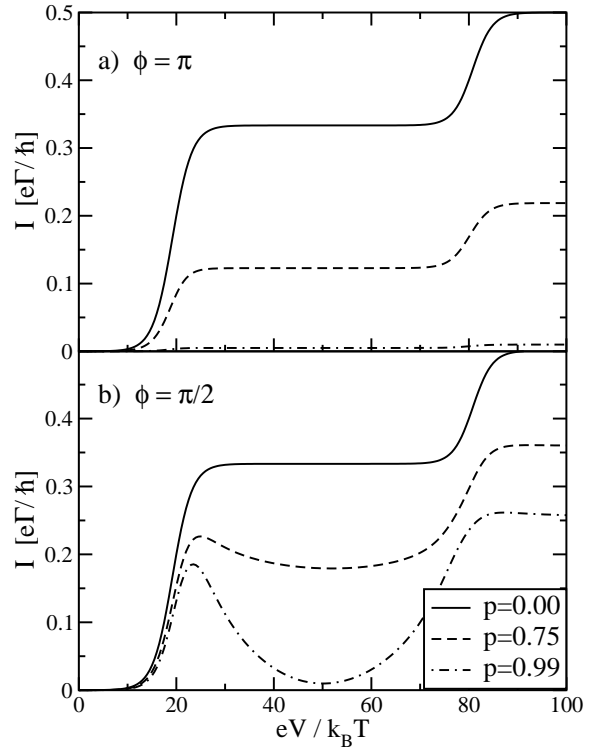}
\caption{\label{currentplots} Current-voltage characteristics for
antiparallel (a) and perpendicular aligned (b) lead
magnetizations. Further
parameters are $\Gamma_{\rm L}=\Gamma_{\rm R}=\Gamma/2$, $p_{\rm L}=p_{\rm R}=p$, $\varepsilon=10{k_{\rm B}T}$, and
$U=30{k_{\rm B}T}.$}
\end{center}
\end{figure}

We now turn to the discussion of the non-linear response regime,
$eV>k_{\rm B}T$.
In Fig.~\ref{currentplots}(a) we show the current $I$ as a function of the bias
voltage $V$ for an antiparallel configuration of the leads' magnetizations
and different values of the leads' spin polarization $p$.
As it is well known from transport through quantum dots with nonmagnetic leads,
the current increases in a steplike manner.
At low bias voltage, the dot is always empty and transport is blocked.
With increasing bias voltage, first single and then double occupancy of the
dot is possible, which opens first one and then two transport channels.
A finite spin polarization $p$ leads to spin accumulation and, thus, to a
reduction of transport.
A reduction of transport with increasing $p$ is also seen for noncollinear
magnetization.
There is a qualitative difference, though.
As can be seen in Fig.~\ref{currentplots}(b), a very pronounced negative
differential conductance evolves out of the middle plateau as the spin
polarization $p$ is increased.

Before explaining the origin of this negative differential conductance,
we address the height of the plateau at large bias voltages first.
Away from the steps, all appearing Fermi functions are either $0$ or $1$.
From Eq.~(\ref{current}) we get $I = (e\Gamma/2\hbar) \left[
1-p{\bm S} \cdot ({\bf \hat n}_{\rm L} - {\bf \hat n}_{\rm R}) \right]$.
When the exchange field is ignored, which can be done here since the spin
dwell time is shortened by the possibility of forming spin singlets, then
the accumulated spin is
${\bm S} = p ({\bf \hat n}_{\rm L} - {\bf \hat n}_{\rm R})/4$
from which we get
\begin{equation}
  I = \frac{e\Gamma}{2\hbar} \left( 1-p^2 \sin^2 \frac{\phi}{2} \right) \, .
\end{equation}
The suppression of transport due to the spin polarization $p$ of the leads is
comparable with the case of a single-tunnel junctions, when charging
effects are of no importance.

At intermediate bias voltages, the dot can be empty or singly occupied,
but double occupation is forbidden.
To understand the negative differential conductance we first neglect the
exchange field and then, in a second step, analyze how the exchange field
modifies the picture.
Since double occupation of the dot is prohibited, all electrons entering the
dot through the left barrier find an empty dot.
This is consistent with the fact that the current
$I = (e\Gamma/\hbar) P_0$ explicitly depends only on the probability to find
the dot empty.
The spin accumulation affects the current only indirectly via modifying $P_0$.
In the absence of the exchange field the relation between them is
\begin{eqnarray}
{\bm S} \!\!&=&\!\! p \left[ \frac{\Gamma_{\rm L}}{\Gamma_{\rm R}}P_0{\bf \hat n}_{\rm L} - \frac{1-P_0}{2}\,{\bf \hat n}_{\rm R} \right] \qquad \text{with}\\
\label{blockade}
  P_0 \!\!&=&\!\! \frac{1}{1+2\Gamma_{\rm L}/\Gamma_{\rm R}}
  \left[ 1+\frac{4}{\Gamma_{\rm R}/\Gamma_{\rm L}+2}\cdot\frac{p^2}{1-p^2}\sin^2 \frac{\phi}{2} \right]^{-1}.
  \quad
\end{eqnarray}
(At this point we explicitly allow for different tunneling strengths.)
As a consequence, with increasing $\phi$ and $p$ the probability $P_0$
decreases, i.e., the dot is most of the time occupied with one electron, and
the accumulated spin tends to align along $-{\bf \hat n}_{\rm R}$,
antiparallel to the drain electrode.
The electron is, thus, trapped in the dot since tunneling out to the
drain electrode is suppressed by a low density of states for the given
spin direction.
No second electron can enter the dot because of the Coulomb interaction,
and transport is, thus, blocked.
So this mechanism is a type of spin blockade but with a different physical
origin compared to the systems described in literature.\cite{spinblock1,spinblock2}

\begin{figure}[!ht]
\begin{center}
\includegraphics[width=0.9\columnwidth,angle=0]{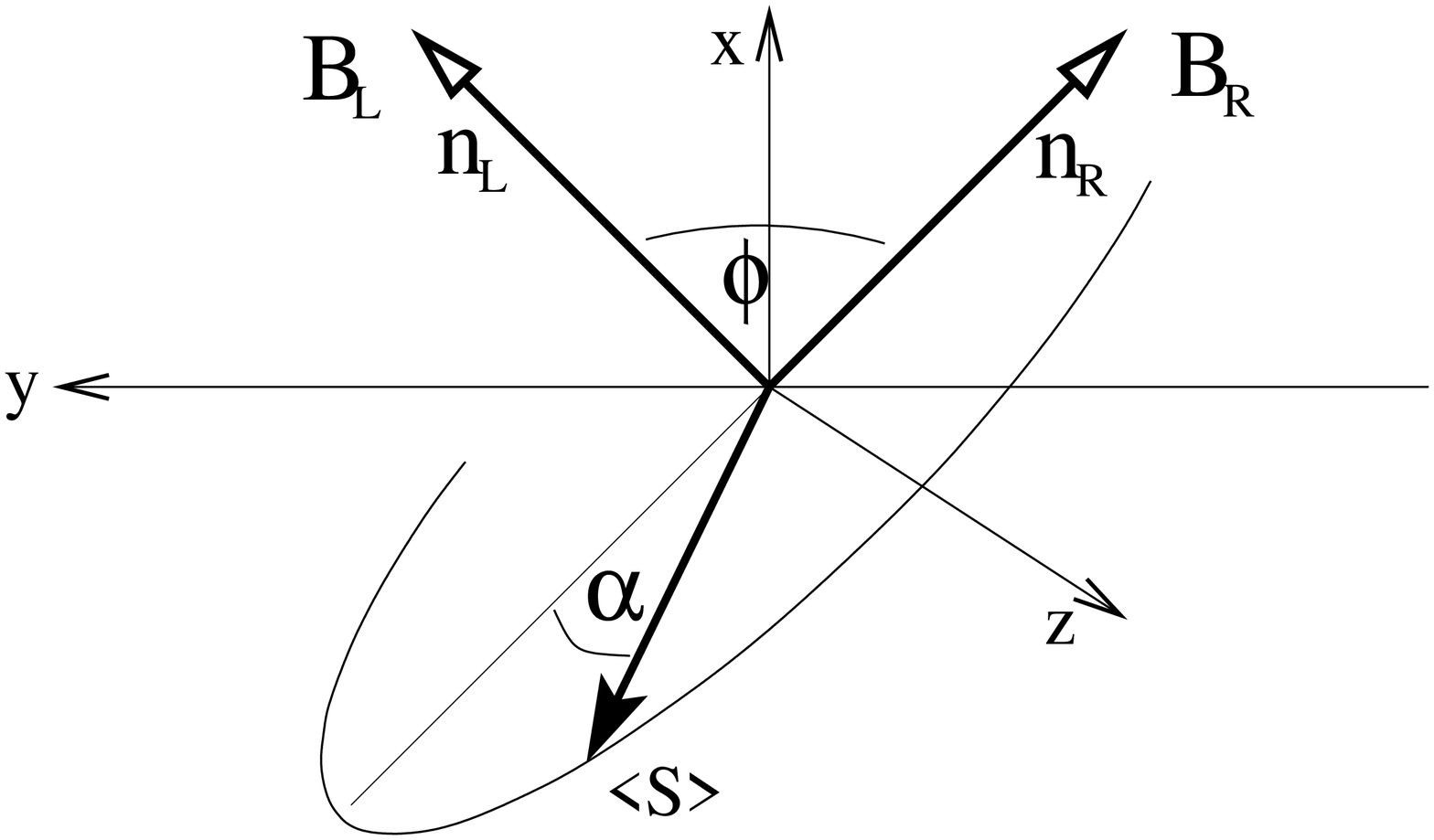}
\caption{\label{dinlr} For electrons polarized antiparallel to
the drain lead, the influence of the effective field generated by
the source lead is dominating. By rotating the spins, the spin
blockade is lifted and therefore the conductance recovers.}
\end{center}
\end{figure}

\begin{figure}[!ht]
\begin{center}
\includegraphics[width=1.0\columnwidth,angle=0]{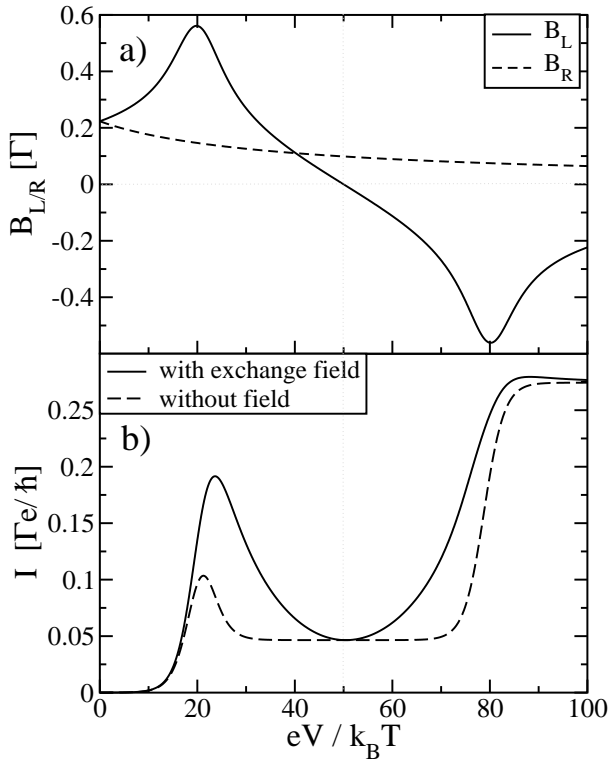}
\caption{\label{exfield} Panel (a) The absolute value of the
effective exchange field contributions from the left and right
leads. Panel (b) the current voltage dependence, with and without
the influence of the exchange field. For both plots the parameters
$\phi=\pi/2$, $\Gamma_L=\Gamma_R=\Gamma/2$,
$\varepsilon=10k_{\rm B}T$, $U=30k_{\rm B}T$, and $p=0.95$ were chosen.}
\end{center}
\end{figure}

The suppression factor coming from Eq.~(\ref{blockade}) defines the local
minimum of the current in Fig.~\ref{currentplots}(b).
At this point, the relevant exchange field component generated by the coupling
to the left lead vanishes, so that spin precession becomes insignificant,
see Fig.~\ref{exfield}(a). Away from this point spin precession sets in as illustrated
in Fig.~\ref{dinlr}. The spin rotates about ${\bf \hat n}_{\rm L}$ and the
electron can now more easily leave the dot via the drain electrode.

Similar to the linear-response regime, the rotation caused by the
exchange field weakens the spin-valve effect.
The two mechanisms, the spin blockade, and the spin precession have opposite
effect on the spin-valve behavior.
Together with the fact that the strength of the exchange field varies as a
function of the level position relative to the Fermi level, we now
fully understand the origin of the nonlinear conductance.
To illustrate this further, we plot in Fig.~\ref{exfield}(b) the current which
we obtain when we drop the contribution, Eq.~(\ref{Srot}), of the exchange field
to the spin dynamics by hand and compare it with the full result.
In the absence of the exchange field, a wide plateau is recovered, whose
height is governed by Eq.~(\ref{blockade}).
The peak at the left end of the plateau indicates that, once the dot level
is close to the Fermi level of the source electrode, the spin blockade is
relaxed since the dot electrons have the possibility to leave to the
left side.

We now estimate under which circumstances the negative differential conductance
will be visible.
Since the spin-blockade mechanism is crucial for the negative differential
conductance to form, the relevant criterion can be obtained from
Eq.~(\ref{blockade}) as
\begin{equation}
  \frac{4}{\Gamma_{\rm R}/\Gamma_{\rm L}+2}\cdot\frac{p^2}{1-p^2}
  \sin^2 \frac{\phi}{2} \sim 1  \, ,
\end{equation}
i.e., for a typical value $\phi \approx \pi/2$ and symmetric coupling
$\Gamma_{\rm L} = \Gamma_{\rm R}$ we need at least a spin polarization
$p \approx 0.77$, in agreement with Fig.~\ref{currentplots}.
An asymmetry in the coupling strength $\Gamma_{\rm L} > \Gamma_{\rm R}$ helps
us to somewhat reduce the number of the required spin polarization.

\begin{figure}[!ht]
\begin{center}
\includegraphics[angle=-90,width=1.0\columnwidth]{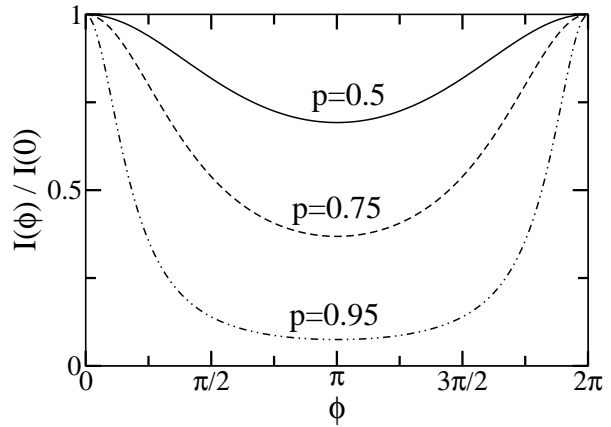}
\caption{\label{cond} Angular dependence of the conductance with
an applied voltage of $V=\varepsilon+U/2$, i.e., the voltage
generating the smallest influence of the exchange field. Further
plot parameters are $\Gamma_{\rm L}=\Gamma_{\rm R}=\Gamma/2$, $p_{\rm L}=p_{\rm R}=p$,
$\varepsilon=10k_{\rm B}T$, and $U=30k_{\rm B}T$.}
\end{center}
\end{figure}

The effect of the spin blockade on the $\phi$-dependent current is depicted
in Fig.~\ref{cond}.
We choose the bias voltage according to $eV/2=\varepsilon+U/2$, such that
the influence of the exchange field is absent.
For $p=0.5$ still a $\sin^2\frac{\phi}{2}$ dependence can be
recognized. For higher polarizations the conductivity drops faster
and stays nearly constant at its minimal value due to spin blockade.
This is just the opposite behavior than predicted for the 
linear-response regime as seen in Fig.~\ref{normalizedconductance}.

We close this section with the remark that while we plotted only results
for the case $\varepsilon >0$, in the opposite case $\varepsilon < 0$ the
current-voltage characteristics is qualitatively the same.

\section{External magnetic field}
\label{sec:magnetic_field}
As we discussed in the previous section, the transport properties through the
quantum dot crucially depend on the magnitude and direction of the average
spin in the dot, which is determined by the interplay of spin-dependent
tunneling and the effective exchange field.
The application of an additional, external magnetic field ${\bm B}_{\rm ext}$
opens the possibility to manipulate the dot spin and, thus, the transport
behavior in a direct way.
Since the external magnetic field is independent of all the system
parameters considered so far, the parameter space for experiments is enlarged
by another dimension.
Magnetic fields have been used either to split the energy levels for the
different spin components\cite{ralph} or to rotate an accumulated
spin.\cite{wees}
A combination of both, a static field splitting the energy levels plus
an oscillating field to rotate the spin, defines an ESR scheme that
allows for the  study of single-spin dynamics in quantum dots in the absence
of ferromagnetic leads, as proposed in Ref.~\onlinecite{ESR}.

The way a static external magnetic field affects the dot qualitatively depends on
the ratio of the induced Zeeman energy and the intrinsic line width $\Gamma$
of the dot energy levels.

\subsection{\label{largefield}Large Fields}
At large magnetic fields, $B_{\rm ext} \gg \Gamma$ (to keep the formulas
transparent, we include the gyromagnetic factor in the definition of
$B_{\rm ext}$), all the complex spin dynamics discussed so far is absent.
It is natural in this case to quantize the dot spin along the direction of
${\bm B}_{\rm ext}$.
The Zeeman splitting term $B_{\rm ext} = \varepsilon_{\uparrow}
- \varepsilon_{\downarrow}$ in the Liouville equation~(\ref{master})
ensures that for weak coupling, the density matrix always remains diagonal,
as can be seen from expanding Eq.~(\ref{master}) to lowest order in $\Gamma$
to obtain $ 0=(\varepsilon_{\uparrow}-\varepsilon_{\downarrow})
{P_{\uparrow}^{\downarrow}}$.
As a consequence, the average spin on the dot is always polarized along the
direction of the external magnetic field, and the exchange field does not play
any role.
In a physical picture, the spin components perpendicular to ${\bm B}_{\rm ext}$
are averaged out due to a fast precession movement.
Transport effectively takes place through two energetically separated levels,
%
%
where the coupling strength to the left and right leads depends on
the angles $\theta_r=\sphericalangle({\bm B}_{\rm ext},{\bf \hat{n}}_r)$
between the lead magnetizations and the external field via the relations
$\Gamma_{r \uparrow}=\Gamma_r(1+p_r\cos\theta_r)$ and
$\Gamma_{r \downarrow} =\Gamma_r(1-p_r\cos\theta_r)$.
Thereby we assumed that the applied magnetic field is not strong enough
to tilt the lead magnetization direction.
It is worth to point out that the spin polarization and the
trigonometric function of the relative angle $\theta_r$ appear
always as a product, so that this system is equivalent to the collinear
cases\cite{QD-theory,ralph} with decreased magnetization of the leads.

\subsection{\label{smallfield}Small Fields}

The situation becomes more interesting in the limit of small fields,
$B_{\rm ext} \lesssim \Gamma$.
To describe this case, we perform the same perturbation expansion in $\Gamma$
as before, but count $B_{\rm ext}$ as being first order in $\Gamma$.
This introduces in Eq.~(\ref{master}) the (first-order) term
$(\varepsilon_{\uparrow}-\varepsilon_{\downarrow})
{P_{\uparrow}^{\downarrow}}$, while the kernels $\Sigma$ are then not
affected by the external magnetic field, because they are already of first
order in the coupling strength $\Gamma$.

It turns out that, in the considered limit, the only change introduced by the
external magnetic field is an additional term ${\bm S}\times{\bm B}_{\rm ext}$
in the Bloch-like equation
\begin{equation}
\label{Bext}
  \frac{d\bm S}{dt} = \left( \frac{d\bm S}{dt}\right)_{\rm acc}
  + \left( \frac{d\bm S}{dt}\right)_{\rm rel} +{\bm S}\times{\bm B}
\end{equation}
with ${\bm B}={\bm B}_{\rm L} +{\bm B}_{\rm R} + {\bm B}_{\rm ext}$.
With the external magnetic field as a new parameter all the effects
of spin blockade and the precession discussed so far can be enhanced or
suppressed and made independent of the lead alignment.

\begin{figure}[!ht]
\begin{center}
\includegraphics[angle=-90,width=1.0\columnwidth]{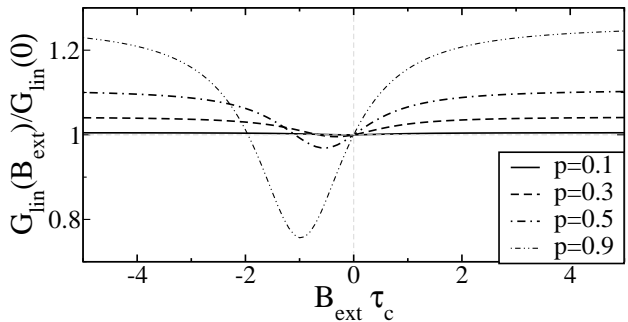}
\caption{Linear conductance as a function of an applied
external magnetic field in $x$ direction. The parameters are
$\Gamma_{\rm L}=\Gamma_{\rm R}=\Gamma/2$, $p_{\rm L}=p_{\rm R}=p$,
$\varepsilon=0$, $U=10k_{\rm B}T$, and $\phi=\pi/2$.
\label{externalfield}
}
\end{center}
\end{figure}

An external field is applied along
${\bf \hat e}_{\rm x} = ({\bf \hat n}_{\rm L} + {\bf \hat n}_{\rm R})/
|{\bf \hat n}_{\rm L}+{\bf \hat n}_{\rm R}|$ leads to a suppression of
the spin-valve effect in the linear-response regime similarly as due to
exchange field.
In fact, according to Eq.~(\ref{Bext}) the magnitude of the relevant field
is just the sum of the exchange and the external fields.
The linear conductance as a function of $B_{\rm ext}$ is displayed in
Fig.~\ref{externalfield}.
In the absence of the effective exchange field the conductance would be
symmetric around $B_{\rm ext} = 0$, and any value of the field would increase
the conductance as discussed in Sec.~\ref{sec:linearresponse}.
The exchange field yields a shift of the position of the minimum.
This opens the possibility to determine the exchange field experimentally
by changing the external magnetic field and keeping all other parameters fixed.
From the generalization of Eq.~(\ref{currlin}) to the present case it becomes
apparent that the width of the minimum is given by the inverse of the dwell
time $\tau_{\rm c}$ in the dot, since during this time the dot spin is
influenced by the fields.

We remark that the effect of external-field components perpendicular to the
$x$ direction is somewhat more complicated, even in the linear-response regime.
The reason is that the induced precession will generate some finite spin
accumulation along the $x$ direction, which introduces a linear correction in
$V$ of the charge occupation probabilities via Eq.~(\ref{P01d}).
This, in turn, acts back on the accumulated spin due to Eq.~(\ref{Sacc}).

\section{Inclusion of spin relaxation}
\label{sec:spin_relaxation}

We emphasize that the only source of spin relaxation considered in our
analysis is the tunnel coupling of the dot states to the left and right leads,
i.e., we require that the spin-flip relaxation time due to other mechanisms such
as coupling to nuclear spins,\cite{hyperfine} or spin-orbit
coupling\cite{SO} is large enough not to destroy spin accumulation.
If we assume a spin relaxation rate $T_2$ equal to the
ensemble spin dephasing time $T_2^{\star}\approx100\,ns$ for n-doped GaAs
bulk material\cite{relax} as lower boundary, it is still justifiable to
neglect spin relaxation effects for interface resistances of the order of
$1M\Omega$.\cite{relax2}
But recent measurements with semiconducting quantum dots indicate even much longer
spin relaxation times.\cite{fujisawa}

Nevertheless, (extrinsic) spin relaxation can be the most limiting factor for
various spin-accumulation experiments in systems of larger
size.\cite{johnson,wees} To include extrinsic spin relaxation
in a simplified phenomenological way, we
add an isotropic relaxation term $-{\bf S}/\tau_{\rm ext}$ in Eq.~(\ref{Srel}),
which leads to
\begin{equation}
\label{blochdamp}
  \left( \frac{d\bm S}{dt}\right)_{\rm rel} =
  -\frac{1}{\tau_{\rm s}}{\bf S}=-\left( \frac{1}{\tau_{\rm c}}+
  \frac{1}{\tau_{\rm ext}} \right){\bf S} \, .
\end{equation}
Now, two characteristic time scales have to be distinguished: the average time
an electron stays in the dot, $\tau_{\rm c}$, and the meanlife time of
a dot spin, $\tau_{\rm s}=(1/\tau_{\rm c}+1/\tau_{\rm ext})^{-1}$.

By solving the system of master equations for the linear-response
regime, we see that we have to replace the factor $\tau_{\rm c}$
with $\tau_{\rm s}$ everywhere in the previous discussions, which matches
the intuition that $\tau_{\rm s}$ is the relevant time scale for spin
related effects.
From Eq.~(\ref{Smag}) we see that the extrinsic relaxation processes reduce the
magnitude of the spin accumulation by a factor $\tau_{\rm s}/\tau_{\rm c}$.
The observation that the accumulated spin on the dot and, therefore, all spin
related effects in transport vanish linearly with
$\tau_{\rm s}/\tau_{\rm c}$ (and not exponentially) implies that
magnetoresistance effects can still be measured, even if the mean dwell time of
the electrons in the dot exceeds the spinlife time significantly,
which was experimentally confirmed by Ref.~\onlinecite{wees}.
The reason for the linear dependence is the following.
Even when the spins relax much faster than the mean dwell time, a fraction of
the spins leaves the dot already before being relaxed.
The probability of detecting a spin with the original alignment in a
transport measurement is given by the time integral
$P_{\rm spin}=\int_{0}^{\infty} dt \exp(-t/\tau_{\rm s})/\tau_{\rm c} =
\tau_{\rm s}/\tau_{\rm c}$, which is linear despite the exponential decay of
the accumulated spin on the dot.

\section{Relation to spin-mixing conductance}
\label{sec:mixing_conductance}

Transport through multiterminal devices with noncollinear lead magnetizations
has been discussed recently by using the language of a complex spin-mixing
conductance $G_{\uparrow\downarrow}$ within a circuit
theory.\cite{spinmix}
The mixing conductance $G_{\uparrow\downarrow}$ contains the information about
the transport of spins oriented perpendicular to the magnetization of the
ferromagnetic leads.
Using this approach, the conductance of a two-terminal
device with noncollinear magnetization of the leads can
be described in linear response by the general relation
\begin{eqnarray}\label{angular_general}
G(\theta)=\frac{G}{2} \left( 1- p^2 \frac{ \tan^2 (\theta/2) }{
\tan^2 (\theta/2) + |\eta |^2/\mathrm{ Re} \, \eta } \right) \; ,
\end{eqnarray}
where $\eta \equiv 2 G_{\uparrow\downarrow}/G $ is the complex
spin-mixing conductance normalized to the sum $G$ of the spin-up and spin-down
conductance of a single contact.
This formula, Eq.~(\ref{angular_general}), is quite general but requires the
knowledge of $\eta$.
Hybrid systems of ferromagnets with normal metals in
diffusive and ballistic regime\cite{spinmix} or even Luttinger
liquids\cite{egger,Balents} have been described.
For system with tunnel contacts $\mathrm{Re} \,\eta = 1$ and Eq.~(\ref{currlin})
can be transformed into the form Eq.~(\ref{angular_general}) by identifying
\begin{eqnarray}
\label{gerri}
  \mathrm{Im}\, \eta \equiv \frac{2 \mathrm {Im} \,
    G_{\uparrow\downarrow}}{G} = B_0\tau_{\rm c} \; .
\end{eqnarray}
Our approach, thus, provides a theory for the interaction effects on the
spin-mixing conductance in the linear-response regime.
The imaginary part $\mathrm{Im}\, \eta$ is proportional to the local effective
field experienced by the dot spins, because this field is responsible for the
spin precession which yields a coherent mixing of the two spin channels.

\section{Conclusions}\label{conclusions}
In the presented work the influence of charging interaction effects on
the tunnel magnetoresistance for systems with noncollinearly magnetized leads
has been discussed.
We examined a single-level quantum connected to ferromagnetic leads as a model
system and found that charge- and spin-related properties do not just coexist
but their interplay gives rise to new effects detectable in transport
measurements.

The most important effects discussed in this paper are related to
the existence of an effective exchange magnetic field
$ {\bm B}_{\rm L} +{\bm B}_{\rm R} $, which leads to a precession movement of
the accumulated spin on the dot.
This precession weakens the spin-valve effect by reducing the amount of
accumulated spin and changing its orientation relative to the leads.
As a result, the current flowing through the system is increased due to the
exchange field.
Also the functional form of the angular-dependent linear conductance is
affected, and the $\cos$-dependence valid for a single magnetic tunnel
junction is modified due to the precession of the accumulated spin.

In nonlinear response, the accumulated spin on the dot tends to
align antiparallel to the drain lead magnetization.
Such a spin blockade causes a suppression of transport for the
bias-voltage regime in which the dot can be empty or singly occupied.
The spin precession due to the coupling to ferromagnetic leads
countersteers the spin blockade by tilting the spin accumulation
direction. Since the strength of the exchange field is a
nonmonotonic function of the applied voltage, a regime with negative
differential conductance exists.

Attaching ferromagnetic leads to quantum dots is quite
challenging. One promising approache is to contact an ultrasmall
aluminum nanoparticle, which serves as a quantum dot, to
ferromagnetic metallic electrodes. In this way, quantum dots with
one magnetic and one nonmagnetic electrode have already been
fabricated.\cite{ralph} An alternative idea is to use carbon
nanotubes as quantum dots and to place them on ferromagnetic
contacts.\cite{cnt} But also other systems are conceivable, such
as self-assembled InAs quantum dots in GaAs LEDs with (Ga,Mn)As as
ferromagnetic electrode,\cite{QD-exp} or STM setups with a natural
or artificial impurity on a ferromagnetic substrate contacted by a
ferromagnetic STM tip.\cite{FMSTM}

\begin{acknowledgments}
We thank J. Barna{\'s}, G. Bauer, A. Brataas, S. Maekawa, G. Sch\"on,
and D. Urban for discussions. This work was supported by the
Deutsche Forschungsgemeinschaft under the Emmy-Noether program and 
the Center for Functional Nanostructures, through SFB491 and GRK726, 
by the EC under the  Spintronics  Network RTN2-2001-00440, and the 
Center of Excellence for Magnetic and Molecular Materials for Future 
Electronics G5MA-CT-2002-04049 and Project PBZ/KBN/ 044/P03/2001.

\end{acknowledgments}

\appendix

\section{\label{diagramtechnique}Diagrammatic Technique}

In this appendix we present some technical details of the derivation of
the master equations and the transport current.
The approach is based on the diagrammatic technique developed in
Ref.~\onlinecite{diagrams}, designed for the application in quantum dot
systems.
The actual calculation is to some degree similar to transport through
Aharonov-Bohm interferometers with quantum dots.\cite{AB2}

\subsection{\label{appendix_sigmas}Transition rates $\Sigma$}

\begin{figure}[!ht]
\includegraphics[width=0.9\columnwidth]{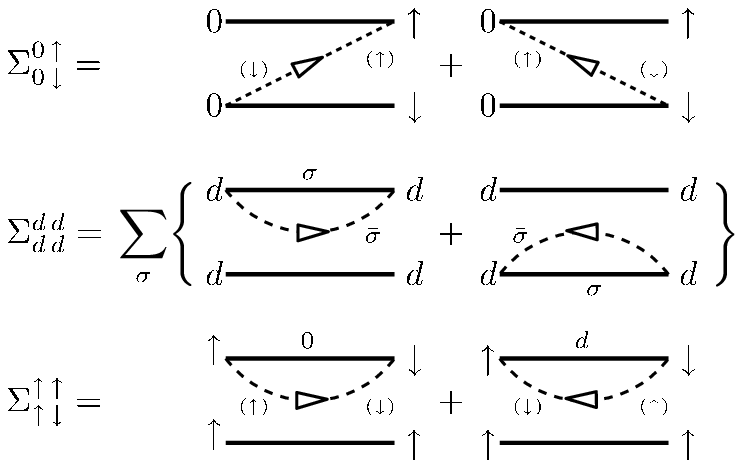}
\caption{\label{dia}Examples for the diagrams representing the
generalized transition rates $\Sigma$.}
\end{figure}

In the Liouville equation (\ref{master}) the transition rates
appear as irreducible self-energies
$\Sigma_{\chi_2^{\prime}\,\chi_2}^{\chi_1^{\prime}\,\chi_1}$.
These self-energy parts are represented as block diagrams enclosed
by a Keldysh contour. Examples of first-order diagrams are shown
in Fig. \ref{dia}. The real-time axis runs horizontally from the
left to the right while the upper (lower) line represents the
forward (backward) propagator of the dot on a Keldysh time contour.
The rules to calculate the $\Sigma$ are:\cite{diagrams}

1.) Draw all topologically different diagrams with tunneling lines connecting
vertices on either the same or opposite propagators.
Assign to the four corners and all propagators the states
$\chi = 0,\uparrow, \downarrow, {\rm d}$ and the corresponding energies
$\varepsilon_{\chi}$, as well as an energy $\omega$ to each tunneling line.

2.) For each time interval on the real axis confined by two
neighboring vertices, assign the resolvent $1/(\Delta E+i0^+)$,
were $\Delta E$ is the energy difference between left and right
going tunnel lines and propagators.

3.) For each vertex connecting a double-occupied state $\rm d$ to the
up state $\uparrow$, the diagram acquires a factor $(-1)$.

4.) Assign to each tunneling line the factor
$\gamma_{\sigma\sigma^{\prime}}^+(\omega)$
[$\gamma_{\sigma\sigma^{\prime}}^-(\omega)$]
when the tunneling line is going backward (forward) with respect to the
Keldysh contour.
Its form is
\begin{eqnarray}
  \gamma^{\pm}_{\sigma\sigma} (\omega) &=&
  \frac{1}{2\pi}\sum_{r, \chi=\pm} \frac{\Gamma_{r\chi}}{2}f_r^{\pm}(\omega)\,,
\\
  \gamma^{\pm}_{\uparrow\downarrow} (\omega)
  &=&\frac{1}{4\pi}\bigl(
  \Gamma_{\rm L+}f_{\rm L}^{\pm}e^{i\phi/2}+
  \Gamma_{\rm R+}f_{\rm R}^{\pm}e^{-i\phi/2}
\nonumber\\
  &&-\Gamma_{\rm L-}f_{\rm L}^{\pm}e^{i\phi/2}
  -\Gamma_{\rm R-}f_{\rm R}^{\pm}e^{-i\phi/2}\,\bigl)
\\ \nonumber
  &=&
  {\gamma^{\pm}_{\downarrow\uparrow}}^{\star} (\omega)\,.\quad
\end{eqnarray}
Here, $\sigma$ and $\sigma^{\prime}$ are the spins of the electron that leaves
and enters the vertices connected by the line.
The factors $\gamma^{\pm}(\omega)$ come from the contraction of two
lead operators in the tunnel Hamiltonian and resemble the transition
rates predicted by Fermis golden rule, including a sum over all
intermediate lead states $(r= {\rm R/L},\sigma=+/-)$. Dependent on the time
ordering, the transition rate is proportional to the electron $f^+$
or the hole distribution function $f^-=1-f^+$, while the relative phases
can be extracted from Fig.~\ref{tunnelphases}.

5.) The diagram gets a prefactor of $(-1)^b(-1)^c$, where $b$ is number of
internal vertices on the backward propagator, and $c$ the number of crossings
of tunnel lines.

6.) Integrate over all energies of the tunneling lines.

In the sequential-tunneling regime, only one tunnel line per
diagram is involved. Therefore, only one frequency
integral appears which can be calculated trivially by using
Cauchy's formula.

\subsection{\label{appendix_green}Green's functions}

For the calculation of the transport current, we need the Fourier transform of
the Keldysh Green's functions,
\begin{eqnarray}
\label{grcalc}
  G^>_{\sigma\sigma^{\prime}}(t) &=&
  -i\aver{\,c^{~}_{\sigma}(t)c^{\dag}_{\sigma^{\prime}}(0)\,}, \\
  G^<_{\sigma\sigma^{\prime}}(t) &=&
  +i\aver{\,c^{\dag}_{\sigma^{\prime}}(0)c^{~}_{\sigma}(t)\,} \, .
\end{eqnarray}
\begin{figure}[!ht]
\begin{center}
\includegraphics[width=0.9\columnwidth,angle=0]{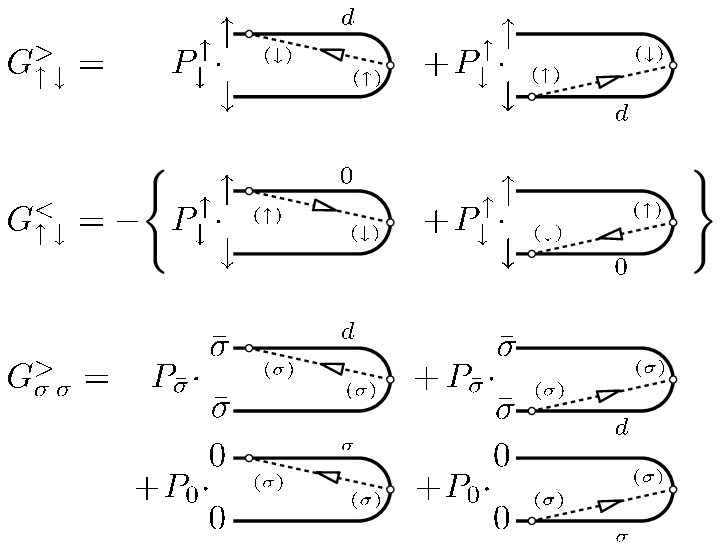}
\caption{\label{gre}Examples for the diagrams representing Green's functions.}
\end{center}
\end{figure}
The Green's function is an average of one creation and
one annihilation operator. These operators appear as external
vertices (open circles) in the diagrams.
They can be constructed following the rules:

1.) Draw the bare contour, the external vertices, and the virtual
tunnel line in the proper orientation as seen in Fig.~\ref{gre}.
The indices of the Green's function indicate the spin of the created and annihilated
electrons at the external vertices.
If higher-order contributions in $\Gamma$ are to be calculated,
add internal vertices according to the desired order and connect them by
tunnel lines.
Multiply the diagram with the matrix element of the density matrix
corresponding to the inital state.
Assign to each propagator with state $\chi$ the corresponding energy
$\varepsilon_{\chi}$ and to each tunnel line and the virtual line an energy
$\omega$.

2.) Apply the same rule as no.(2) for calculating $\Sigma$.

3.) Apply the same rule as no.(3) for calculating $\Sigma$.

4.) Assign each tunneling line the factor
$(-1)^v\gamma_{\sigma\sigma^{\prime}}^+(\omega)$, where $v$ is the
number of external vertices on the Keldysh contour enclosed by the
end points of the tunnel line. The line connecting the external
vertices does not contribute.

5.) Apply the same rule as no.(5) for calculating $\Sigma$.

6.) Integrate over all energies except the energy assigned
to the virtual tunnel line.

For the lesser function $G^<_{\sigma\sigma^{\prime}}$, the direction
of the external tunnel lines is reversed, and an additional global
minus sign is involved in the definition.

In this paper we assume weak dot-lead coupling, i.e., we need only the
zeroth-order Green's functions. In this limit, we get
$G^{>}_{\sigma\sigma} (\omega) =-2 \pi i P_{\bar{\sigma}} \delta(\omega - \varepsilon - U)
-2 \pi i P_{0} \delta(\omega - \varepsilon)$,
$G^{<}_{\sigma\sigma} (\omega) =+2 \pi i P_{\sigma} \delta(\omega - \varepsilon)
+2 \pi i P_{d} \delta(\omega - \varepsilon - U)$ and for $\sigma \neq \bar{\sigma}$ we find
$G^{>}_{\sigma\bar{\sigma}} (\omega) =2 \pi i P^{\sigma}_{\bar{\sigma}} \delta(\omega - \varepsilon -U)$,
$G^{<}_{\sigma\bar{\sigma}} (\omega) =2 \pi i P^{\sigma}_{\bar{\sigma}} \delta(\omega - \varepsilon)$.


\begin{thebibliography}{99}
\bibitem{remark}
In the opposite limit, i.e., strong dot-lead coupling, Kondo correlations
may arise.
Their interplay with ferromagnetism in the leads has been discussed in
Refs.~\onlinecite{Kondo}.
\bibitem{com1}
We emphasize that this absence of spin accumulation is a peculiarity of the
weak-coupling limit.
Corrections of higher order in the coupling strength $\Gamma$ will yield a
net spin accumulation due to the spin-dependent energy level renormalization
discussed in the previous section, but this is beyond the scope of this paper.
\end{thebibliography}
\end{document}